\documentclass[conference]{IEEEtran}
\IEEEoverridecommandlockouts

\usepackage{cite}
\usepackage{amsmath,amssymb,amsfonts}
\usepackage{algorithm}
\usepackage{algpseudocode}
\usepackage{graphicx}
\usepackage{textcomp}
\usepackage{multirow}
\usepackage{multicol}
\usepackage{hyperref}
\usepackage{xcolor}
\usepackage{booktabs}
\usepackage{subcaption}
\def\BibTeX{{\rm B\kern-.05em{\sc i\kern-.025em b}\kern-.08em
    T\kern-.1667em\lower.7ex\hbox{E}\kern-.125emX}}
\begin{document}

\title{Differentially Private Federated Quantum Learning via Quantum Noise}
\author{\IEEEauthorblockN{
Atit Pokharel,
Ratun Rahman,
Shaba Shaon,
Thomas Morris,
Dinh C. Nguyen}
\IEEEauthorblockA{{Department of Electrical and Computer Engineering},
{The University of Alabama in Huntsville}, USA
}}

\maketitle

\begin{abstract}

Quantum federated learning (QFL) enables collaborative training of quantum machine learning (QML) models across distributed quantum devices without raw data exchange. However, QFL remains vulnerable to adversarial attacks, where shared QML model updates can be exploited to undermine information privacy. In the context of noisy intermediate-scale quantum (NISQ) devices, a key question arises: How can inherent quantum noise be leveraged to enforce differential privacy (DP) and protect model information during training and communication? This paper explores a novel DP mechanism that harnesses quantum noise to safeguard quantum models throughout the QFL process. By tuning noise variance through measurement shots and depolarizing channel strength, our approach achieves desired DP levels tailored to NISQ constraints. Simulations demonstrate the framework’s effectiveness by examining the relationship between differential privacy budget and noise parameters, as well as the trade-off between security and training accuracy. Additionally, we demonstrate the framework's robustness against an adversarial attack designed to compromise model performance using adversarial examples, with evaluations based on critical metrics such as accuracy on adversarial examples, confidence scores for correct predictions, and attack success rates. The results reveal a tunable trade-off between privacy and robustness, providing an efficient solution for secure QFL on NISQ devices with significant potential for reliable quantum computing applications.

\end{abstract}


\section{Introduction}
Quantum Federated Learning (QFL) is an innovative framework where quantum devices in a network train a shared model using local quantum neural networks (QNNs) and local models exchange without sharing the raw data. Grounded by the principles of federated learning (FL), QFL eliminates the need to centralize potentially sensitive data\cite{huang2022quantum, pokharel2025quantum, pokharel2024electrocardiogram}. The growing interest in QFL is driven by its capability to utilize quantum advantages, such as enhanced model capacity through high-dimensional Hilbert spaces and faster convergence on complex tasks in a distributed training environment \cite{araujo2024quantum}. With ongoing quantum hardware advancements and their increasing availability \cite{abughanem2025ibm}, QFL applications have become pragmatic and scalable in the noisy intermediate-scale quantum (NISQ) era.

Despite a promisingly secure architecture for the raw data, QFL is not immune to security challenges posed by adversarial attacks in a sparse network of quantum devices \cite{montalbano2025quantum}, as model updates shared over the network can allow adversaries to infer sensitive information. Additional risks arise at inference time, where attackers can exploit the trained global model through adversarial queries \cite{10633422}. These test-time threats are often overlooked in QFL settings, posing a significant security challenge. Such vulnerabilities highlight the need for comprehensive defense mechanisms in QFL systems.

Differential privacy (DP) \cite{wei2020federated} presents a well-established solution to these security concerns by introducing noise to mask individual data contributions and further enhancing robustness during inference by acting as a form of regularization. The noise-induced smoothing effect with a reduction in the model’s sensitivity makes it more resistant to adversarial examples. In classical machine learning, DP is typically implemented by adding artificial noise to the training process. However, in the context of QFL with NISQ devices, which are inherently noisy, a compelling question arises: \textit{Can quantum noise in NISQ devices provide differential privacy in QFL network?}

\textbf{Main Contributions:} In this paper, we affirmatively answer this question by introducing differentially-private QFL (DP-QFL), a novel framework that capitalizes inherent noise in NISQ devices as the primary mechanism for enhancing security via DP in a QFL network. The key contributions of this paper are summarized as follows: 
\begin{itemize}
    \item This work introduces the DP-QFL framework to address security in quantum federated networks where DP arises from the combined effect of shot noise and inherent depolarizing noise, with measurement shots and the depolarizing channel strength adjusting the overall noise variance to balance privacy guarantees and NISQ hardware limitations. \textit{To the best of our knowledge, this work is the first to study DP in a QFL system with inherent quantum noise.}
    \item{We implement a quantum-based  adversarial attack following the approach in \cite{akter2024quantum, lu2020quantum} to simulate a realistic threat scenario. The robustness of the proposed DP-QFL framework is then evaluated against this attack using a range of performance and resilience metrics.}
    \item{With extensive simulations, we assess the training performance of DP-QFL under varying privacy budgets, achieved by different quantum noise levels, using two baseline datasets for comparison.}
\end{itemize}

\begin{figure}[ht!]
    \centering
    \includegraphics[width=\linewidth]{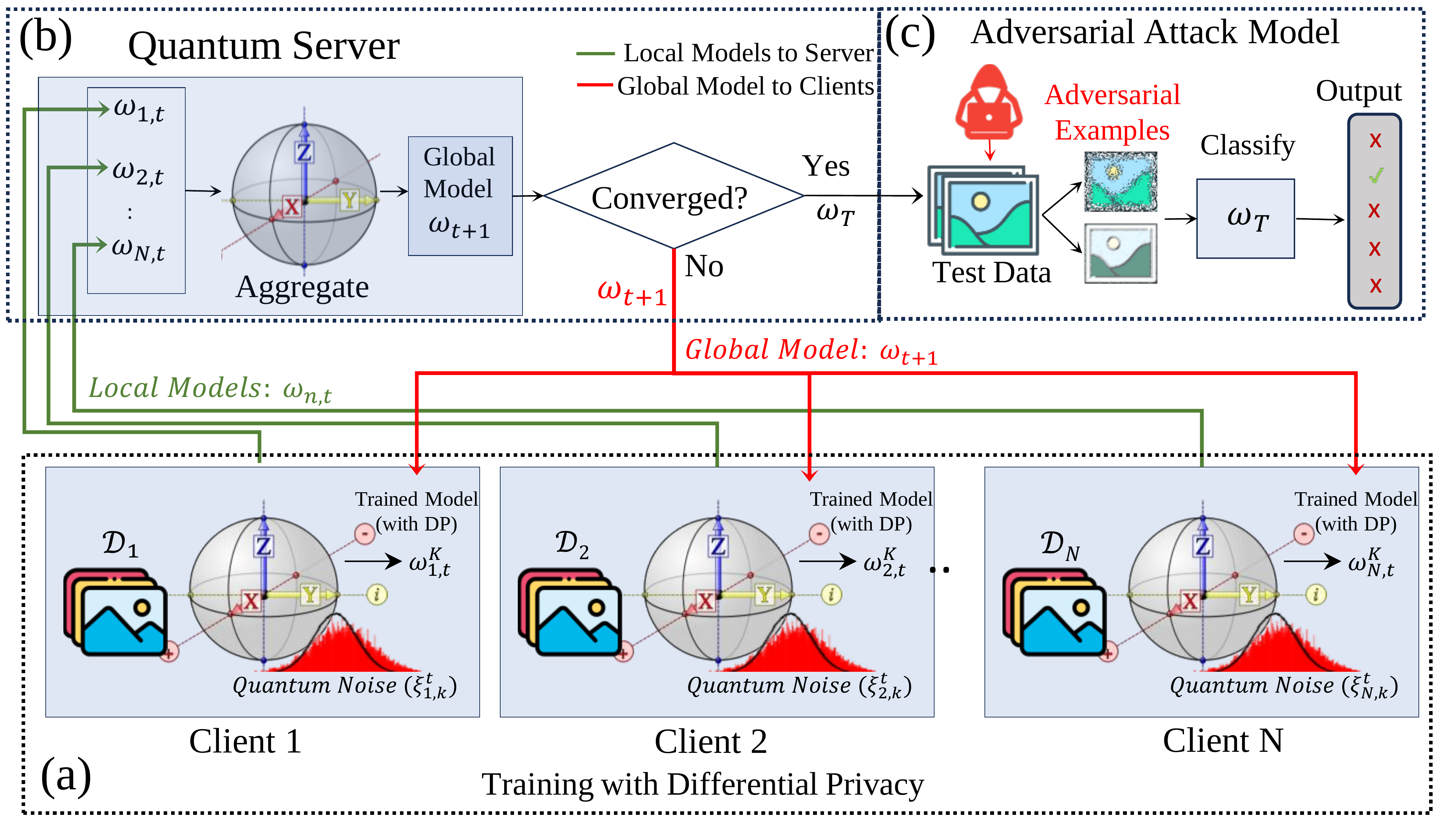} 
    \caption{Overview of the DP-QFL framework. (a) NISQ devices as clients train their local model with a DP mechanism using inherent noise and update the server (\textit{green lines}). (b) The quantum server aggregates the models from the clients and sends back an aggregated global model (\textit{red lines}) for further training. (c) An adversary creates adversarial examples on the test data, which is used to evaluate the fully trained global model.}
    \label{fig:qfl_combined}
\end{figure}
Fig. \ref{fig:qfl_combined} provides an overview of the DP-QFL framework. In Fig. \ref{fig:qfl_combined}(a), NISQ devices act as clients, each training a local model with differential privacy via quantum noise. The locally updated models are sent to the quantum server, as indicated by the green lines. Fig. \ref{fig:qfl_combined}(b) shows the quantum server aggregating the received models to form a global model, which is then distributed back to the clients for continued training, indicated by the red lines. In Fig. \ref{fig:qfl_combined}(c), an adversary generates adversarial examples on the test data to evaluate the fully trained global model.

\section{Related Works}
We summarize related works from various aspects relevant to our work.

\noindent
\textit{1. Quantum Federated Learning:}
QFL has drawn increasing interest in recent years for enabling collaborative training across a distributed network of quantum devices without exchanging raw data \cite{rahman2025sporadic}. A significant number of prior works \cite{qiao2024transitioning, larasati2022quantum, park2025entanglement,gurung2024personalized} have developed innovative QFL algorithms to facilitate the distributive training in a network of quantum devices, while other works have focused on introducing QFL in a diverse set of applications like healthcare \cite{bhatia2024communication}, satellite systems \cite{park2024dynamic}, intrusion detection \cite{abou2024privacy}, load prediction \cite{rahman2025electrical}, environments \cite{rahman2025multimodal} etc. A set of literature \cite{wei2024hybrid, subramanian2024hybrid} also discusses hybrid classical-quantum systems integrated into the FL framework for better performance in their specific tasks. Although these QFL frameworks preserve data privacy via data locality in training quantum devices, the risk of adversarial attacks is still prevalent and undermined in the literature.

\noindent
\textit{2. Quantum Noise as Security Mechanism:}
Despite the fact that quantum noise can degrade the training performance in quantum machine learning (QML) frameworks, several studies have leveraged its inherent randomness to enhance model security. Depolarizing noise was shown to enforce differential privacy by \cite{Bai_2024, yang2023improved}, while \cite{hirche2023quantum} demonstrated its role in reducing information leakage from an information theory perspective. Multiple sources of quantum noise are utilized to enforce the differential privacy in \cite{ju2024harnessing, du2021quantum, zhou2017differential}, where the noise sources are tuned to achieve the desired DP budget. These mechanisms are discussed for standalone QML algorithms where sharing of raw data to a centralized training quantum device is required in the first place. Applying DP in a privacy-preserving QFL framework requires additional considerations tailored to it.

\noindent
\textit{3. DP in Quantum Federated Learning:}There have been very few attempts to enhance the security of a QFL system with a DP mechanism. Authors in \cite{rofougaran2024federated} introduced DP in the QFL scenario with the injection of artificial noise inspired by the classical DP mechanisms. Another similar work \cite{ullah2024quantum} used Laplacian noise to provide DP in their QFL algorithm. The capability of quantum random number generators was used to generate noise that ensures DP in QFL was discussed in \cite{gupta2024enhancing}. Although these efforts somehow attempt to enhance the security of the QFL system, the potential of quantum inherent noises as a mechanism to guarantee privacy is barely discussed.

Thus, we aim to fill these research gaps in this work by proposing a novel DP-QFL framework where the unavoidable inherent noise in NISQ devices is used to enhance the security of a QFL system.

\section{Proposed Methodology}
\subsection{Data Encoding and Local Model Training}
QFL adapts federated learning principles to quantum systems. It enables \( N \) distributed clients to collaboratively train a shared model while preserving data privacy. In QFL, each client at global round \( t \) and local epoch \( k \) encodes classical input data \( w \in \mathbb{R}^d \) into a quantum state using an encoding unitary \( U_{\text{enc}}(w) \). The resulting encoded state is given by
\begin{equation}\label{eq:encoding}
|\psi_{\text{enc}}(w)\rangle = U_{\text{enc}}(w) |0\rangle^{\otimes D},
\end{equation}

where \( D \) is the number of qubits. This encoded state is processed by a parameterized quantum circuit (PQC), represented as
\begin{equation}\label{eq:pqc output}
|\psi_{\text{out}}(w, \omega_{n,t}^{k})\rangle = U(\omega_{n,t}^{k})|\psi_{\text{enc}}(w)\rangle.
\end{equation}

The PQC itself is composed of \( L \) layers of parameterized rotational gates. The total unitary operation on PQC $U(\omega_{n,t}^{k})$ is given by $ \prod_{d=1}^L U_d(\omega_{n,t,d}^{k})$
,where each gate layer $U_d(\omega_{n,t,d}^{k})$ is defined as $\exp\left(-i \frac{\omega_{n,t,d}^{k}}{2} G_d \right)$, and \( G_d \) is a Pauli-string generator.
To evaluate model performance, the output state is measured using a Hermitian observable \( O \), which in our case is the Pauli-Z operator. The measured observable output is given by in an ideal case is given by

\begin{equation}\label{eq:output loss}
f(w, \omega_{n,t}^{k}) = \langle \psi_{\text{out}}(w, \omega_{n,t}^{k}) | O | \psi_{\text{out}}(w, \omega_{n,t}^{k}) \rangle. 
\end{equation}

In practice, this value is estimated through $M$ measurement shots, giving the empirical output estimate $\hat{f}_{n,t}^{k}(\omega)$ as $\frac{1}{M} \sum_{j=1}^{M} H_j$, 
where \( H_j \in \{-1, +1\}\) denotes individual measurement outcomes. The local loss function for each client is defined as:
\begin{equation}\label{eq:loss function}
L_{n,t}^{k}(\omega_{n,t}^{k}) = \frac{1}{|\mathcal{D}_{n,t}|} \sum_{(w, y) \in \mathcal{D}_{n,t}} \ell(y, \hat{f}_{n,t}^{k}(\omega_{n,t}^{k})).
\end{equation}

Here, \( \ell \) is the chosen loss function (e.g., categorical cross-entropy), \( y \) is the label for the input data point and \(|\mathcal{D}_{n,t}|\) is the dataset size.

\subsection{Noise Models}
We consider quantum noise intrinsic to NISQ devices, focusing on two primary types: shot noise arising from measurements and depolarizing noise introduced by quantum gates. We examine the combined effects of these noise sources during parameter updates.

\textbf{Shot Noise:} Shot noise arises from finite \(M\) measurements. For each component, the variance of noise can be represented as
\begin{equation}\label{eq:shot noise variance}
\text{Var}(\hat{f}_{n,t}^{k}) = \frac{\text{Tr}(O^2)}{M},
\end{equation}

since we are considering the Pauli-Z operator, \(O^2 = I\), and \(\text{Tr}(I) = 2^D\) for \(D\) qubits. Thus, the variance $\text{Var}(\hat{f}_{n,t}^{k})$ becomes $ \frac{2^D}{M}$. According to \cite{ju2024harnessing}, individual measurements yield a binomial distribution (\(\pm 1\) outcomes) but for large \(M\), \(\hat{f}_{n,t}^{k}\) approximates a Gaussian via the Central Limit Theorem (CLT), providing the foundation for the combined noise’s Gaussian nature.

\textbf{Depolarizing Noise:} As highlighted in studies of quantum noise, depolarizing noise captures gate imperfections across quantum circuits \cite{du2021quantum}, transforming the output state as
\begin{equation}\label{eq:depol noise}
    \rho_E = (1 - p) |\psi_{\text{out}}\rangle\langle\psi_{\text{out}}| + p \frac{I}{2^D},
\end{equation} 
where $p$ defined by $1 - \prod_{i=1}^{L} (1 - p_i)$ represents the cumulative error probability with uniform per-gate error \(\lambda = p_i\) across \(L\) layers, further simplified as $1 - (1 - \lambda)^L$. A similar formulation in \cite{zhou2017differential} emphasizes this channel’s prevalence in quantum devices, mixing the ideal state with the maximally mixed state \(\frac{I}{2^D}\). The resulting noisy expectation, as derived in these works for an observable \(O\) like Pauli-Z, is \(f_E = (1 - p) f(w, \omega_{n,t}^{k})\), since \(\text{Tr}(O \frac{I}{2^D}) = 0\), introducing a scaling to the output by \(1 - p\) \cite{zhou2017differential}. According to \cite{zhu2021bias}, such data-independent bias that is uniform across inputs $w$ does not affect the DP guarantee, as it satisfies invariance under post-processing. 

The variance of measurement outcomes, as quantified by \cite{jose2022error,du2021quantum}, increases with \(c(p) = 1 - (1 - p)^2\), reflecting the binomial nature of shot noise. A study in \cite{ju2024harnessing} asserts that while depolarizing noise itself isn’t Gaussian, the circuit’s total noise output can be approximated as Gaussian under averaging over many shots, a claim rooted in Lindeberg's central limit theorem \cite{linnik1959information}. Thus, since the depolarizing noise introduces a data-independent bias, we focus on its scaling effect on the overall noise variance in the gradient. In this work, we approximate the total noise as Gaussian, with a uniform, data-independent scaling of the measurement variance by \(c(p)\), determined by the cumulative depolarizing probability. Also, we always consider this noise at least at the minimum value, approximating the hardware error rate to account for the unavoidable limit. If required by the QFL-DP algorithm, the depolarizing channel can be intentionally made more noisy, but never reduced below this intrinsic threshold.

\subsection{Noisy Gradient Estimation}
Noise from quantum measurements propagates into the gradient estimation process. This occurs because the gradient is derived from noisy observable measurements, introducing variance directly into the gradient components.
We define the \emph{noisy} gradient for parameter dimension \(d\) using the parameter-shift rule \cite{wang2022qoc} expressed as
\begin{equation}\label{eq:param shift}
[\hat{g}_{n,t}^{k}]_d 
= \tfrac{1}{2}\Bigl(\hat{f}_{n,t}^{k}\bigl(\omega_{n,t}^{k} + \tfrac{\pi}{2}e_d\bigr) 
    - \hat{f}_{n,t}^{k}\bigl(\omega_{n,t}^{k} - \tfrac{\pi}{2}e_d\bigr)\Bigr),
\end{equation}
where \(\hat{f}_{n,t}^{k}\) is the empirical measurement output, and \(e_d\) is the unit vector along \(d\). 
Shot noise (\(\tfrac{1}{M}\)) and depolarizing noise (\(c(p)\)) together yield, for a single data point \(w\), variance is
\begin{equation}
\mathrm{Var}\bigl([\hat{g}(w)]_d\bigr) 
= \tfrac{2^D\,c(p)}{2\,M}.
\end{equation}
Averaging over the local dataset \(\!\mathcal{D}_{n,t}\) (size \(|\mathcal{D}_{n,t}|\)) introduces an additional factor from \(\tfrac{\partial\ell}{\partial f}\), assumed \(\le B\). Denoting the noise as 
\(\xi_{n,t}^{k} = \hat{g}_{n,t}^{k} - \nabla L_{n,t}^{k}\), total noise variance can be represented as
\begin{equation}\label{eq:total_noise_variance}
\mathrm{Var}\bigl([\xi_{n,t}^{k}]_d\bigr) 
\;\le\; \tfrac{2^D\,B^{2}\,c(p)}{2\,M\,\lvert \mathcal{D}_{n,t}\rvert}.
\end{equation}
Averaging over \(|\mathcal{D}_{n,t}|\) data points and a sufficiently large \(M\) shots exploits shot noise to bound the variance, governing the noise amplitude. In contrast, gate depolarizing noise, through \(c(p)\), scales the variance randomly and independently with the underlying data, which supports the variance control essential for the quantum differential privacy framework. Parameters are updated over \(K\) local epochs with learning rate \(\eta\), yielding the total update \(\omega_{n,t}^{K} - \omega_{n,t}^{0} = - \eta \sum_{k=0}^{K-1} \hat{g}_{n,t}^{k}\). The total noise \(\xi_{n,t}^{\text{total}} = - \eta \sum_{k=0}^{K-1} \xi_{n,t}^{k}\) has variance per component given by
\begin{equation}\label{eq:variance per component}
\text{Var}([\xi_{n,t}^{\text{total}}]_d) = K \eta \cdot \mathrm{Var}\bigl([\xi_{n,t}^{k}]_d\bigr) ,
\end{equation}
where the scaling by \(K\) and \(\eta\) reflects the accumulation of per-epoch noise contributions.

\subsection{Global Aggregation in QFL}

After $K$ local epochs, each client $n$ obtains an updated parameter vector
$\omega_{n,t}^{K}$. These updates are sent to the central server, which
aggregates them into a global model $\omega_{t+1}$ for the next round, which can be represented as
\begin{equation}\label{eq:global aggregation}
\omega_{t+1}
\,=\,
\frac{1}{N}\,\sum_{n=1}^{N}\,\omega_{n,t}^{K}.
\end{equation}
This averaging rule ensures that the global model reflects contributions from all participating clients while maintaining decentralized data privacy. The process is repeated for \(T\) rounds, allowing the server to iteratively refine the global model based on locally computed noisy gradients.

\subsection{Differential Privacy Mechanism}
Let \(\rho = \rho(\mathcal{D}_{n,t})\) and \(\rho' = \rho(\mathcal{D}_{n,t}')\) be two quantum states encoded for neighboring datasets differing by one data point \((w, y)\). A randomized mechanism \(\mathcal{M}\) satisfies \((\epsilon, \delta)\)-differential privacy \cite{zhou2017differential} if, for all measurable sets \(S \subseteq \mathbb{R}^{{D}_\omega}\),
\begin{equation}\label{eq:DP definition}
\Pr[\mathcal{M}(\rho) \in S] \le e^{\epsilon} \Pr[\mathcal{M}(\rho') \in S] + \delta.
\end{equation}
Here, \(\mathcal{M}(\rho)\) denotes the classical parameter update \(\omega_{n,t}^{K} - \omega_{n,t}^{0}\).
 Thus, with gradient clipping \(\|\hat{g}_{n,t}^{k}\|_2 \le C\), the sensitivity $\Delta_{n,t}$ is formulated as
 $\eta K \frac{2C}{|\mathcal{D}_{n,t}|}.$ 
To satisfy \((\epsilon_{n,t}, \delta)\)-DP, we require
\begin{equation}\label{eq: epsilon nt}
\epsilon_{n,t} = \frac{\Delta_{n,t}}{\sigma} \sqrt{2 \ln\left(\frac{1.25}{\delta}\right)},
\end{equation}
where $\sigma = \sqrt{\text{Var}([\xi_{n,t}^{\text{total}}]_d)}$ is the standard deviation corresponding to its total variance.

For \(N\) clients over \(T\) communication rounds, per-client per-round \((\epsilon_{n,t}, \delta)\)-DP composes via advanced composition theorem \cite{kairouz2015composition}, expressed as
\begin{equation}\label{eq:advanced composition}
\epsilon_{\mathrm{total}}
\,=\, \sqrt{2NT \,\ln\!\Bigl(\tfrac{1}{\delta'}\Bigr)}\,\epsilon_{n,t}
\;+\;\tfrac{NT\,\epsilon_{n,t}\,\bigl(e^{\epsilon_{n,t}} - 1\bigr)}{2},  
\end{equation}
where $\delta_{\mathrm{total}}.
\,=\, NT\,\delta \;+\;\delta'.$
This improves over naive \((NT)\,\epsilon_{n,t}\) by treating privacy loss as a random variable. The first term in \eqref{eq:advanced composition} leverages Chernoff bounds for sublinear growth, while the latter term corrects non-linear accumulation (approximating \(\tfrac{NT\,\epsilon_{n,t}^2}{2}\) for small \(\epsilon_{n,t}\)). Consequently, \(\bigl(\epsilon_{\mathrm{total}}, \delta_{\mathrm{total}}\bigr)\)-DP across QFL is tighter.

\subsection{Proposed  Algorithm}

\begin{algorithm}[!ht]
\footnotesize
\caption{Differentially Private Quantum Federated Learning (DP-QFL)}
\label{algo:dpqfl}
\begin{algorithmic}[1]
\State \textbf{Input:} Global rounds \( T \), local epochs \( K \), clients \( \mathcal{N} \), learning rate \( \eta \), shots \( M \).
\State \textbf{Initialize:} Global model \( \boldsymbol{\omega}_{0} \).
\For{each round \( t = 0, \dots, T{-}1 \)}
    \State Server sends \( \boldsymbol{\omega}_{t} \) to all clients.
    \For{each client \( n \in \mathcal{N} \) in parallel}
        \State Update client models \( \omega_{n,t}^{0} \gets \boldsymbol{\omega}_{t} \).
        \For{each local epoch \( k = 0, \dots, K{-}1 \)}
            \State Encode input using amplitude encoding via Eq. \eqref{eq:encoding}.
            \State Apply PQC with depth \( L \), depol noise rate \( \lambda \) \eqref{eq:depol noise}.
            \State Estimate noisy gradient via parameter shift via Eq.\eqref{eq:param shift}.
            \State Clip gradients to bound sensitivity.
            \State Update local model, \( \omega_{n,t}^{k+1} \gets \omega_{n,t}^{k} - \eta \hat{g}_{n,t}^{k} \).
        \EndFor
        \State Client sends noisy update \( \omega_{n,t}^{K} \) to server.
    \EndFor
    \State Server aggregates \( \boldsymbol{\omega}_{t+1} = \frac{1}{N} \sum_n \omega_{n,t}^{K} \).
\EndFor
\State \textbf{Output:} Final model \( \boldsymbol{\omega}_{T} \) with \((\epsilon_{\text{total}}, \delta)\)-DP via Eq.~\eqref{eq:advanced composition}.
\end{algorithmic}
\end{algorithm}

Algorithm~\ref{algo:dpqfl} summarizes the DP-QFL process. The server initializes the global model (line~2) and runs \(T\) communication rounds (line~3). In each round, clients perform local training with noisy PQC-based updates (lines~6--11), clip gradients (line~10), and send results to the server (line~12). The server aggregates updates (line~13). DP is enforced through inherent quantum noise and advanced composition (line~15).

\section{Simulations and Evaluation}

\begin{figure}[ht!]
    \centering
    \includegraphics[width=\linewidth]{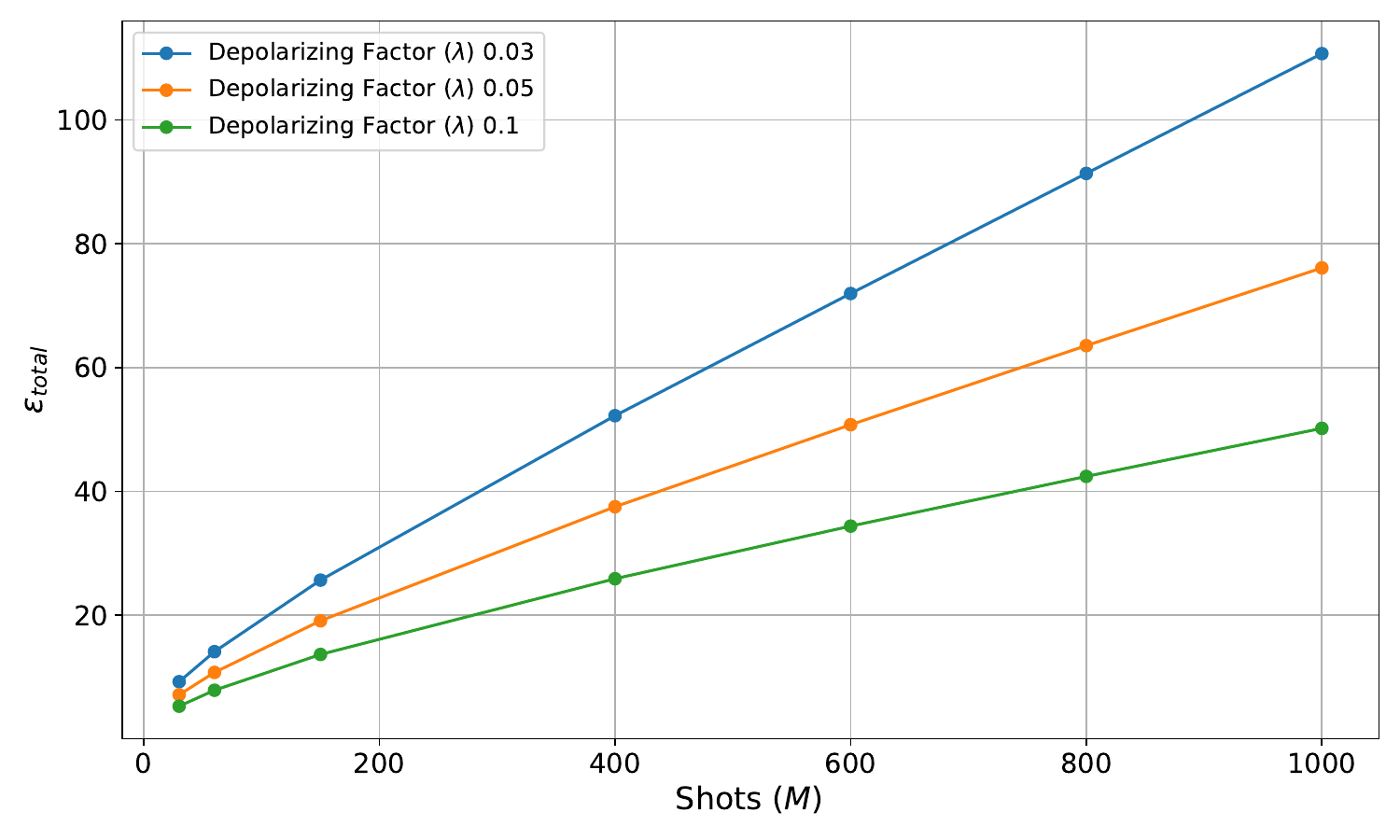} 
    \caption{Relation of privacy budget in DP-QFL with the varying number of shots and depolarizing factor. }
    \label{fig:shots_vs_epsilon}
\end{figure}
\subsection{Dataset and Simulation Setting}
We evaluate our DP-QFL framework on two widely used benchmark datasets: MNIST and CIFAR-10, processed independently in separate experiments. The MNIST dataset consists of 60,000 training and 10,000 test images of handwritten digits (0-9). Each image is in grayscale with a dimension of 28$\times$28 pixels. The CIFAR-10 dataset includes 50,000 training images and 10,000 test images across 10 object classes, each with a dimension of 32$\times$32 pixels. The dimensionality of the images is reduced to 16$\times$16 to facilitate quantum encoding using a limited number of qubits. 

Simulations were performed on a server equipped with an NVIDIA GeForce RTX 4090 GPU, 64GB RAM, and running Ubuntu 22.04. A state vector simulator with the Torchquantum library was used for quantum circuit modeling, integrated with PyTorch for classical machine learning components. 

To map the image feature vectors into their respective quantum states, amplitude encoding is used. The PQC consists of $L$ layers, with each layer including 3 rotational gates ($R_x, R_y, R_z$) per qubit, followed by CNOT entangling gates between adjacent qubits. For the federated learning setup, 10 clients are simulated, with each conducting 5 local training epochs with a batch size of 64. The global model was updated over 50 communication rounds. Training was optimized using the Adam optimizer with a learning rate of 0.01, minimizing the cross-entropy loss function. To ensure DP by bounding the sensitivity of the gradients, we applied a gradient clipping threshold of 0.8. This mitigates the impact of outlier gradients and distinctive features, improving privacy guarantees.
\subsection{DP-QFL Performance Evaluation}
\begin{figure*}[hbt!]
    \centering
    \footnotesize
    \begin{subfigure}[t]{0.24\textwidth}
        \centering
        \includegraphics[width=\linewidth]{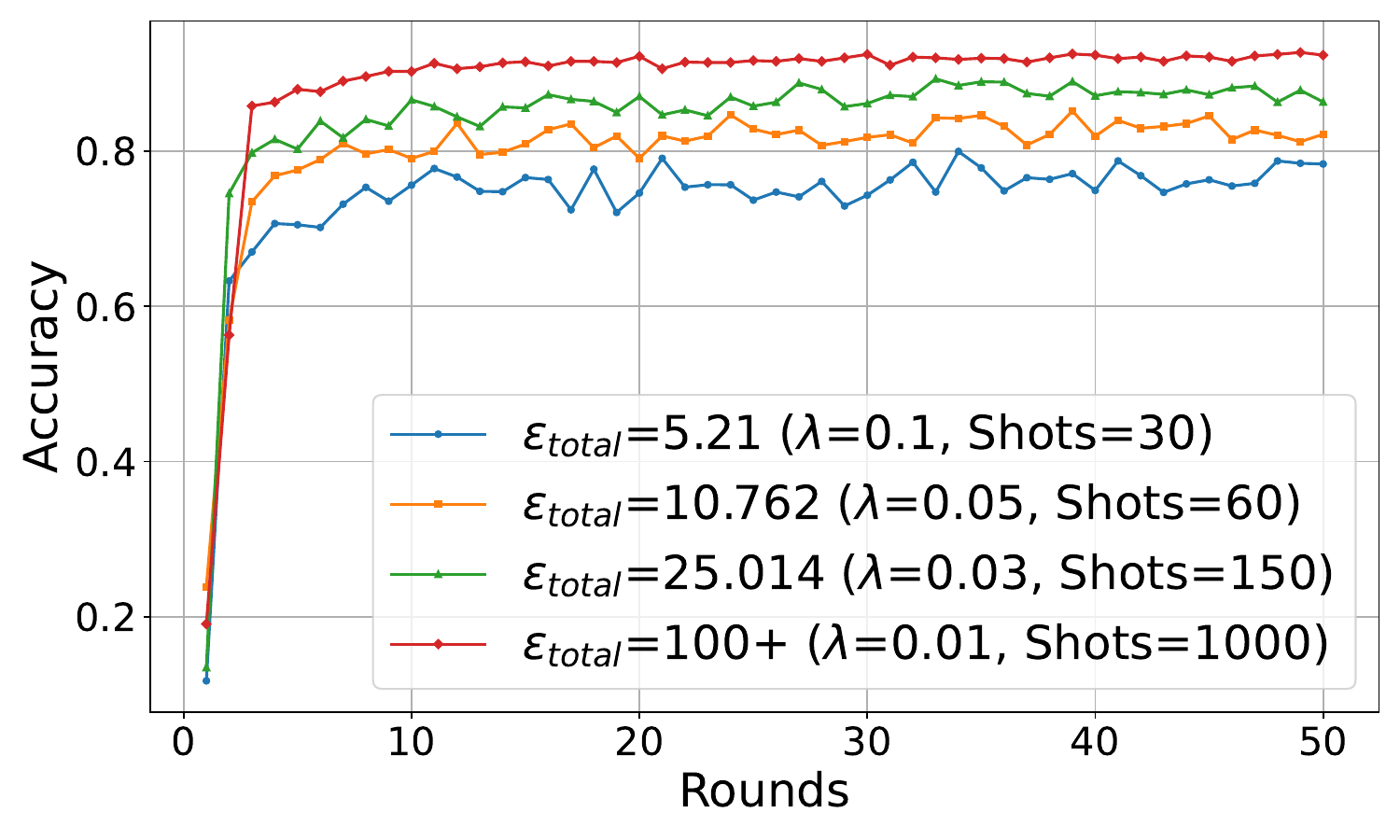}
        \caption{\footnotesize Accuracy on MNIST}
        \label{fig:train_acc_mnist}
    \end{subfigure}
    \hfill
    \begin{subfigure}[t]{0.24\textwidth}
        \centering
        \includegraphics[width=\linewidth]{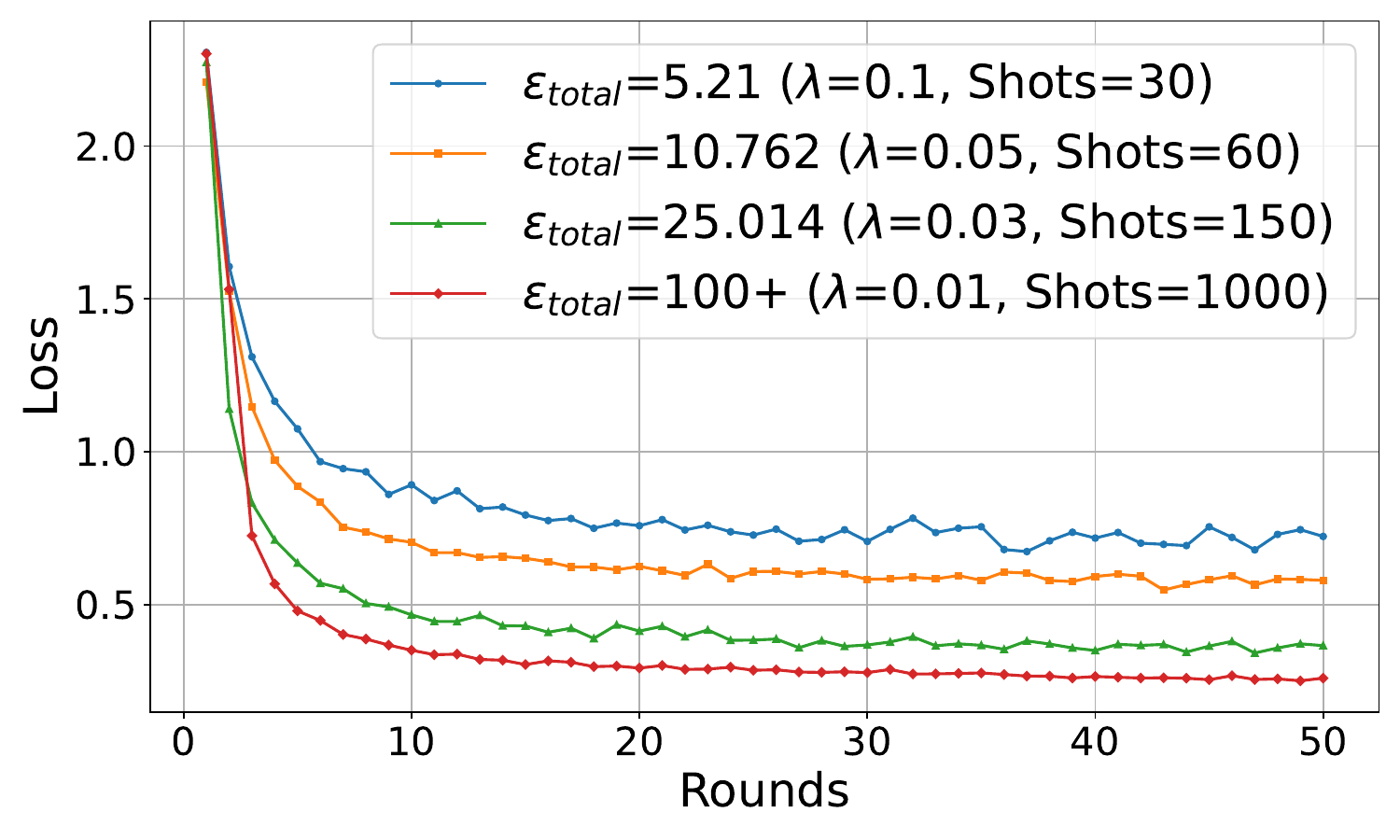}
        \caption{\footnotesize Loss on MNIST}
        \label{fig:train_loss_mnist}
    \end{subfigure}
    \hfill
    \begin{subfigure}[t]{0.24\textwidth}
        \centering
        \includegraphics[width=\linewidth]{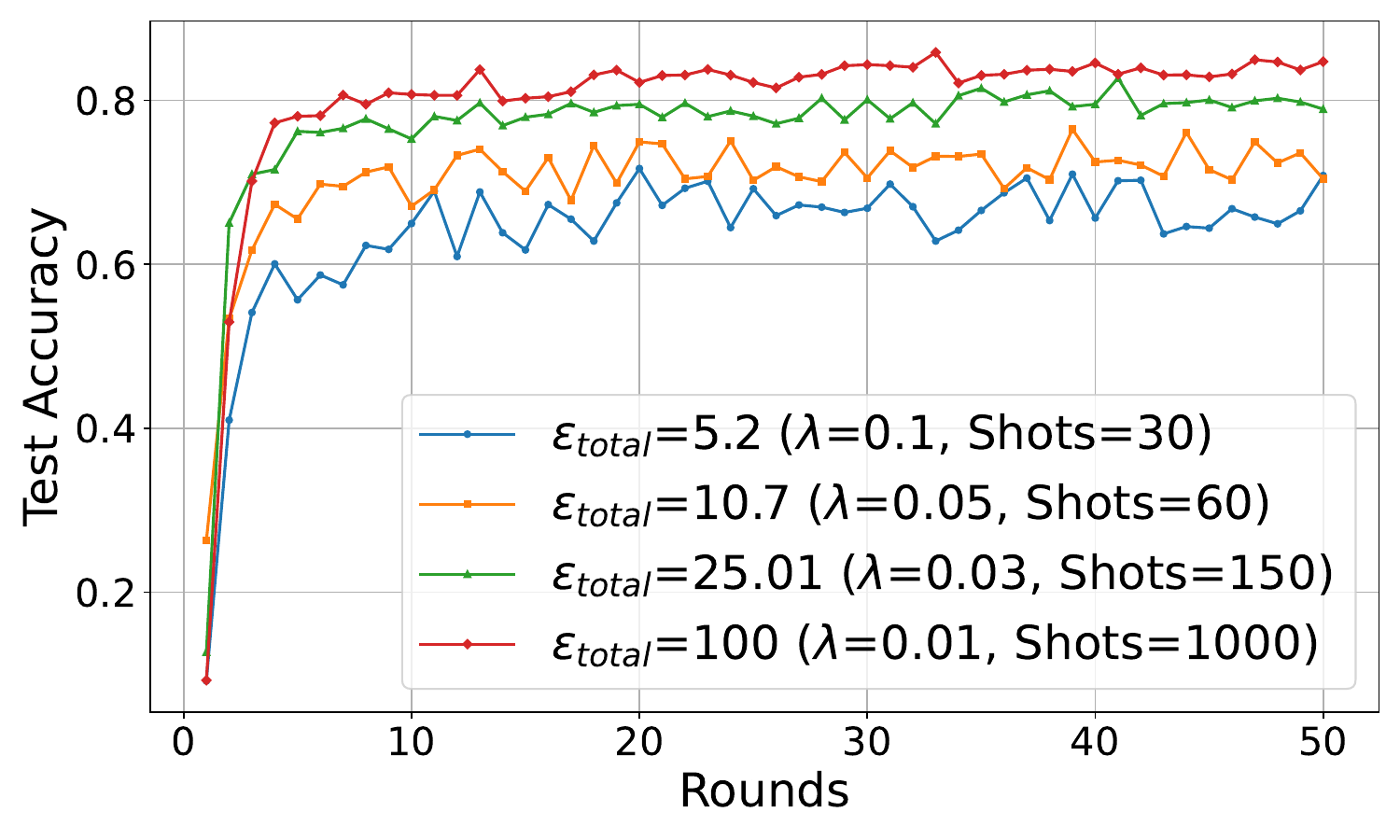}
        \caption{\footnotesize Accuracy on CIFAR-10}
        \label{fig:test_acc_cifar}
    \end{subfigure}
    \hfill
    \begin{subfigure}[t]{0.24\textwidth}
        \centering
        \includegraphics[width=\linewidth]{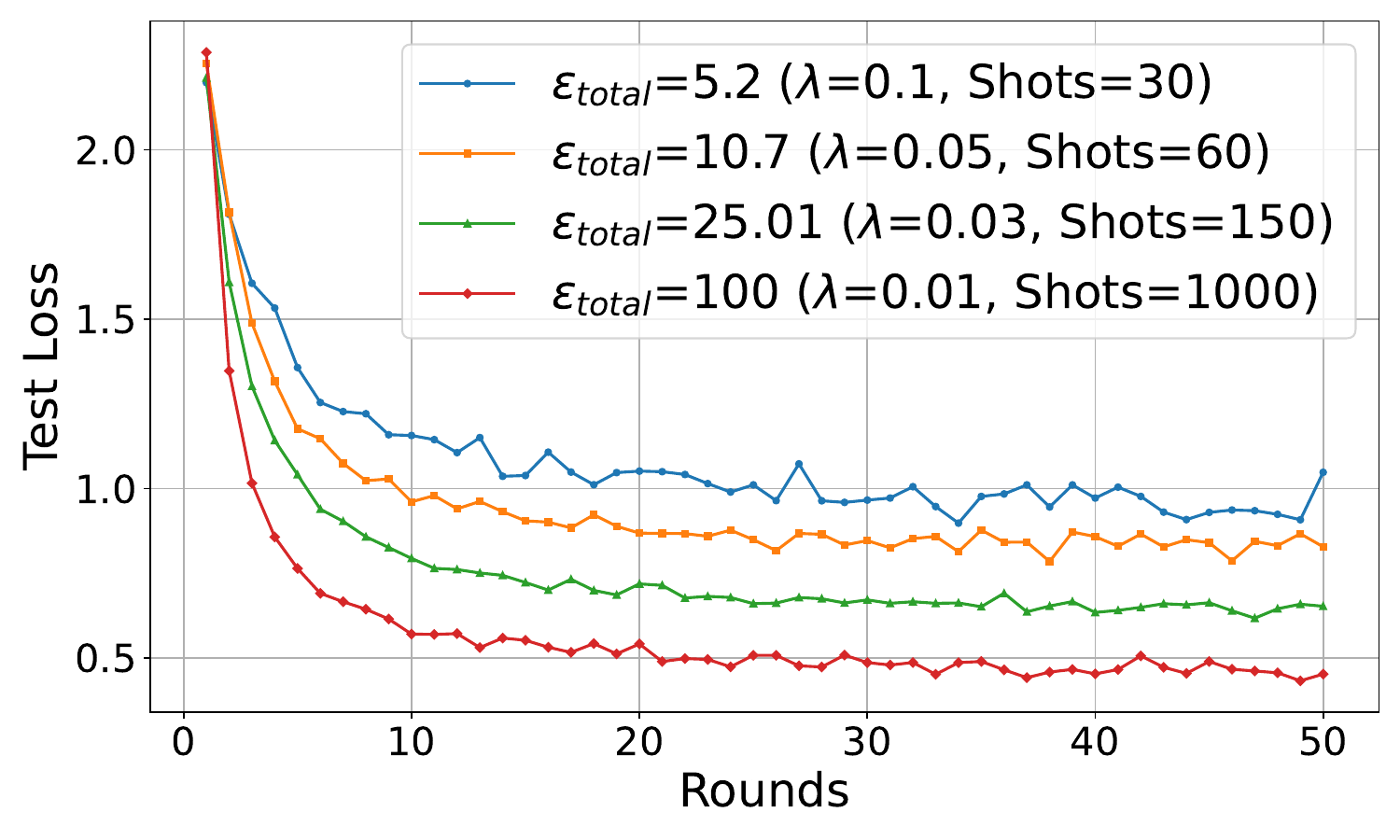}
        \caption{\footnotesize Loss on CIFAR-10}
        \label{fig:test_loss_cifar}
    \end{subfigure}

    \vspace{2pt}
    \caption{Training accuracy and loss curves of DP-QFL with MNIST and CIFAR-10 datasets.}
    \label{fig:qfl-1x4-all}
    \vspace{-2mm}
\end{figure*}
\textbf{Privacy budget analysis: }We assess the performance of the DP-QFL framework by analyzing the relation between DP guarantees and quantum noise parameters, specifically the number of measurement shots $M$ and depolarizing factor $\lambda$ for the PQCs configuration with 8 qubits and 3 layers. Fig. \ref{fig:shots_vs_epsilon} illustrates how the noise sources contribute to achieving the desired privacy budget ($\epsilon_{total}, \delta_{total}$) in the DP-QFL setting. It is noticed that the privacy budget decreases with a higher number of shots due to the corresponding increase in statistical noise variance, as depicted in equation \eqref{eq:total_noise_variance}. Similarly, it can also be inferred that a lower depolarizing noise factor leads to reduced total noise variance, resulting in a higher privacy budget. Thus, both increased shots and lower depolarization make the system less private by lowering the inherent noise. In our simulation, the strongest privacy guarantee with an $\epsilon_{total}$ of approximately 5$\pm$0.1 is achieved with just 30 shots and a depolarizing factor of 0.01. In contrast, with shots over 1000 and a lower depolarizing factor, the privacy budget exceeds 100, indicating that the DP-QFL provides minimal security in such settings. In all simulations, a fixed $\delta_{total} = 10^{-5}$ ensures a tight privacy bound, consistent with the advanced composition theorem given in equation \eqref{eq:advanced composition}. This simulation allows us to conclude that a good $\epsilon_{total}$ can be attained by tuning primarily the number shots, aided by the contribution of inherent depolarizing noise in the total noise variance.

\textbf{Training Performance of DP-QFL: }The training performance under varying privacy budgets is evaluated through accuracy and loss curves, as depicted in Fig. \ref{fig:qfl-1x4-all}. For the MNIST dataset, figures \ref{fig:train_acc_mnist} and \ref{fig:train_loss_mnist} show that the least secure model with a privacy budget over 100 has an accuracy of over 94\%. Meanwhile, the most secure version in our simulation with a privacy budget of $5\pm0.2$ had a trade-off with performance, achieving an accuracy of 79\%. A similar privacy-performance trade-off is observed for the CIFAR-10 dataset, as shown in Fig.~\ref{fig:test_acc_cifar} and Fig. ~\ref{fig:test_loss_cifar}. The highest accuracy of 86\% is achieved under minimal privacy protection, using 1000 measurement shots and a depolarizing factor of 0.01. As the total privacy budget $\epsilon_{total}$ is reduced to 25.01, 10.7, and 5.2, the corresponding accuracies decrease to 81\%, 76\%, and 71\%, respectively, highlighting the inverse relationship between privacy strength and model performance.
The high-privacy setting, with lower $\epsilon_{total}$, corresponds to the standard noise levels found in NISQ quantum devices. In contrast, low-noise settings that lead to higher privacy budgets require very high shot counts and low gate noise, which are not practical with current hardware. This makes strong privacy not just achievable but also a natural fit for real-world quantum systems.

\begin{table}[htbp]
\caption{Performance of DP-QFL compared to the state-of-art methods.}
\centering
\footnotesize
\begin{tabular}{lcccc}
\toprule
\multirow{2}{*}{{Loss Function}} & \multicolumn{2}{c}{{MNIST}} & \multicolumn{2}{c}{{CIFAR-10}} \\
\cmidrule(lr){2-3} \cmidrule(lr){4-5}
& Accuracy & Loss & Accuracy & Loss \\
\midrule
QML-DP \cite{yang2023improved, du2021quantum} & {84.54\%} & {0.56\%} & {73.22\%} & {0.93\%} \\
QFL-AN \cite{rofougaran2024federated, ullah2024quantum}          & 83.01\% & 0.61\% & 75.88\% & 0.88\% \\
DP-QFL (Ours)          & \textbf{87.60\%} & \textbf{0.37\%} & \textbf{79.9\%} & \textbf{0.66\%} \\
\bottomrule
\end{tabular}
\label{tab:comp_loss}
\end{table}

It is essential to provide a comparative analysis between our DP-QFL model and existing state-of-the-art approaches, which include standalone quantum machine learning (QML) models and differentially private QFL methods based on artificial noise injection. Table~\ref{tab:comp_loss} summarizes the performance of these models in comparison to our proposed DP-QFL framework. Firstly, we test the standalone QML performance with the DP (QML-DP) privacy budget formulation discussed in \cite{yang2023improved} and \cite{du2021quantum}. For the same level of security, QFL-DP surpasses standalone systems in test results due to collaborative learning across distributed clients, which results in better generalization. Secondly, we compare our framework to a closely related approach \cite{rofougaran2024federated, ullah2024quantum} that incorporates differential privacy into QFL through the injection of artificial noise (QFL-AN), such as Gaussian or Laplacian mechanisms. For this simulation, we use a minimal amount of unavoidable quantum noise reflecting the NISQ scenario. With the same amount of theoretical privacy budget, QFL-DP surpasses these approaches, as shown in the test results. This is mainly due to the additional effect of external noise injected, getting compounded with inherent quantum noise, resulting in degraded performance.
\begin{table}[ht]
\caption{Effect of varying number of qubits ($D_q$) on performance and privacy in DP-QFL.}
\centering
\footnotesize
\begin{tabular}{lccc}
\toprule
\multirow{1}{*}{{Qubits $D_q$}} & \multicolumn{1}{c}{{MNIST Acc. (\%)}} & \multicolumn{1}{c}{{CIFAR-10 Acc. (\%)}} & \multicolumn{1}{c}{$\epsilon_{total}$} \\
\midrule
8   & \textbf{86.18} & 78.02 & 10.762 \\
10  & 85.88 & \textbf{78.61} & 4.700 \\
12  & 80.47 & 72.19 & \textbf{2.290} \\
\bottomrule
\end{tabular}
\label{tab:varying_qubits}
\vspace{-2mm}
\end{table}
The effect of circuit depth of PQCs and varying qubit count on the training performance, along with the privacy, was analyzed. For this experiment, the number of $M$ and $\lambda$ were set to 60 and 0.05, respectively.
Table~\ref{tab:varying_qubits} shows the effect of increasing qubit count on model performance and privacy. While more qubits reduce $\epsilon_{total}$ due to higher inherent noise, they also increase circuit complexity. At 10 qubits, accuracy is high and $\epsilon_{total}$ is strong, showing a good trade-off. But at 12 qubits, performance drops, likely from overfitting and excess noise, indicating that too many qubits can harm generalization despite better privacy.
\begin{table}[ht]
\caption{Effect of varying number of PQC layers ($l$) on performance and privacy in DP-QFL.}
\centering
\footnotesize
\begin{tabular}{lccc}
\toprule
\multirow{1}{*}{{Layers $l$}} & \multicolumn{1}{c}{{MNIST Acc. (\%)}} & \multicolumn{1}{c}{{CIFAR-10 Acc. (\%)}} & \multicolumn{1}{c}{$\epsilon_{total}$} \\
\midrule
3 & \textbf{87.37} & \textbf{79.30} & 12.001 \\
4 & 86.18 & 78.02 & 10.762 \\
5 & 87.00 & 78.54 & \textbf{9.022} \\
\bottomrule
\end{tabular}
\label{tab:varying_layers}

\end{table}
Table~\ref{tab:varying_layers} shows how increasing PQC depth affects accuracy and privacy. More layers improve privacy by increasing circuit noise, reducing $\epsilon_{total}$. However, performance already peaks at 3 layers. Beyond this, deeper circuits add optimization difficulty without clear accuracy gains, showing that shallow circuits can achieve better trade-offs in practice.

\subsection{Adversarial Attack on QFL}
\textbf{Attack model:}
To evaluate the robustness of our DP-QFL framework, we perform an analysis of our model's performance under a quantum-based black-box adversarial attack model discussed in \cite{lu2020quantum} and \cite{akter2024quantum}. In this setting, the adversary doesn't have access to the global model's parameters, circuit structure, or training data. Instead, the attacker can only query the global model by submitting an input \( x_{\text{adv}} \in \mathbb{R}^d \) and receives an output prediction \( f(x_{\text{adv}}, \omega_t) \).

\textbf{Attack mechanism:} A synthetic labeled dataset \( D_{\text{sub}} \) is generated by querying the victim model (trained DP-QFL) with inputs $x_{\text{adv}}$ and recording the corresponding outputs. The adversary constructs a quantum substitute model parameterized by \( \omega_{\text{sub}} \). The substitute QNN is then trained using cross-entropy loss 
$
L_{\text{sub}}(\omega_{\text{sub}}).
$
Subsequently, the adversary computes the input gradient \( \nabla_w L_{\text{sub}} \) using the parameter-shift rule on the substitute model to approximate how small perturbations affect the loss. 
Using the estimated gradient, the adversary generates an adversarial input as
\begin{equation}\label{eq:adv attack}
x_{\text{attack}} = x_{\text{adv}} + E \cdot \text{sign}(\nabla_w L_{\text{sub}})
\end{equation}
where E is the attack noise strength. 
\begin{figure}
    \centering
    \includegraphics[width=\linewidth]{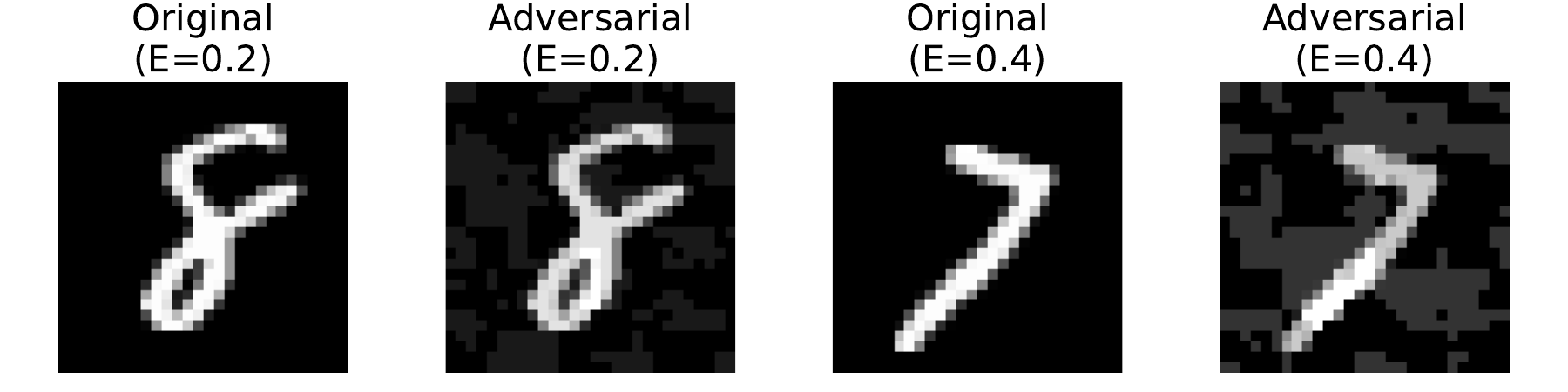} 
    \caption{Adversarial examples generated using a black-box QNN with carefully crafted noise that appears indistinguishable yet causes the model to misclassify.}
    \label{fig:adv_examples}
    \vspace{-3mm}
\end{figure}
Fig.~\ref{fig:adv_examples} is the visualization of adversarially modified MNIST samples under varying noise strengths \( E \in [0.1, 0.4] \). Although the true characteristics of the digits are preserved, visually retaining class-identifiable features, the injected perturbations induce subtle pixel-level distortions. These distortions, visible at higher \( E \) if observed closely, remain indistinguishable to humans but are sufficient enough to generate targeted misclassifications in the DP-QFL model. Such an attack model poses a major threat in critical service applications, where modified inputs to deployed models can lead to catastrophic failures.

\textbf{DP-QFL Robustness Evaluation:}  We evaluate the robustness of DP-QFL under black-box fast gradient sign method (FGSM) attacks using adversarial examples generated with increasing noise strength $E$ ranging from $0.12$ to $0.20$. Two global models are considered: one trained under a moderate privacy budget to mimic a secure model($\epsilon_{\text{total}} = 5.2$) and another less secure model with minimal DP noise ($\epsilon_{\text{total}} > 100$). For convenience, we denote the minimally secure model as the non-DP model and the secure model as the DP-protected model in this analysis.

\begin{figure}
    \centering
    \label{fig:attack}
    \footnotesize
    \begin{subfigure}[t]{0.32\linewidth} 
        \centering
        \includegraphics[width=\linewidth]{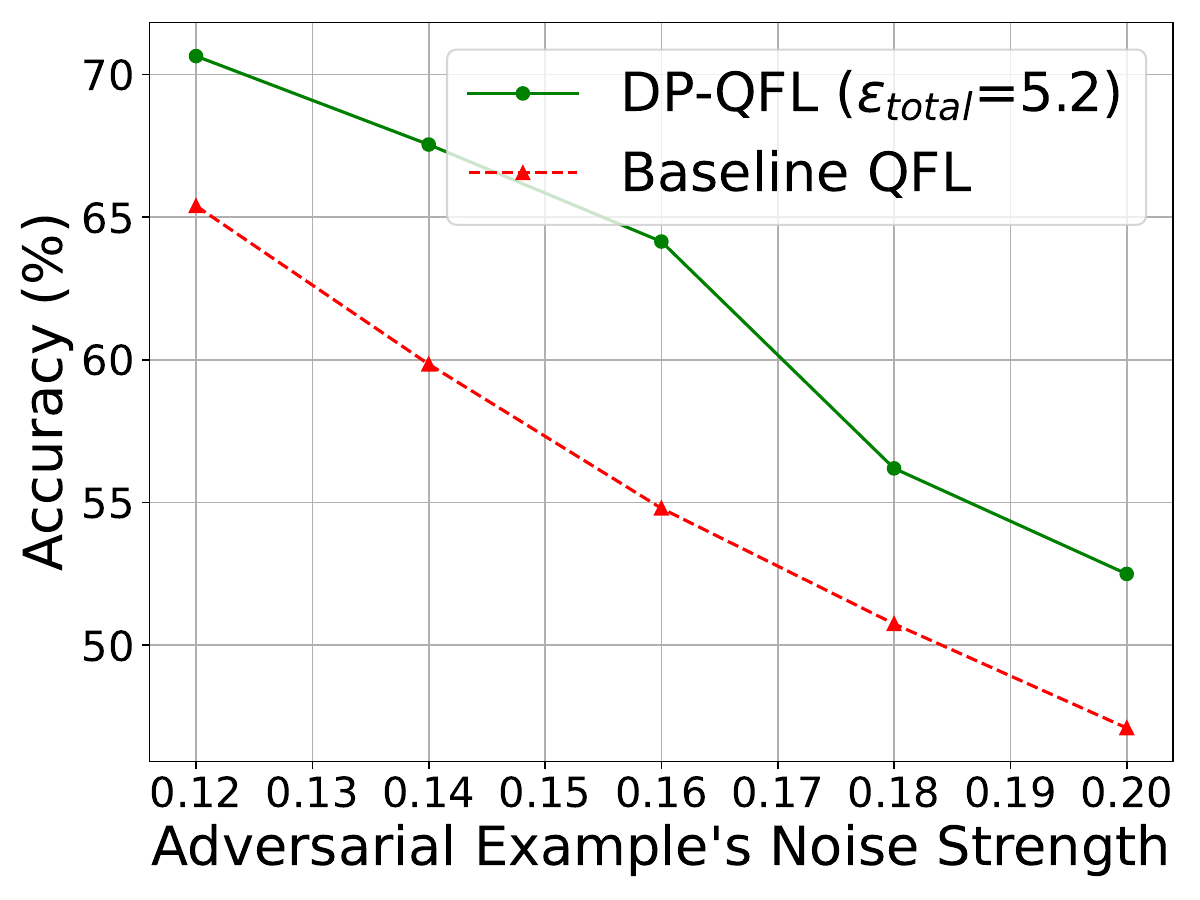}
        \caption{\footnotesize Adversarial classification accuracy}
        \label{fig:adversarial accuracy}
    \end{subfigure}
    \begin{subfigure}[t]{0.32\linewidth}
        \centering
        \includegraphics[width=\linewidth]{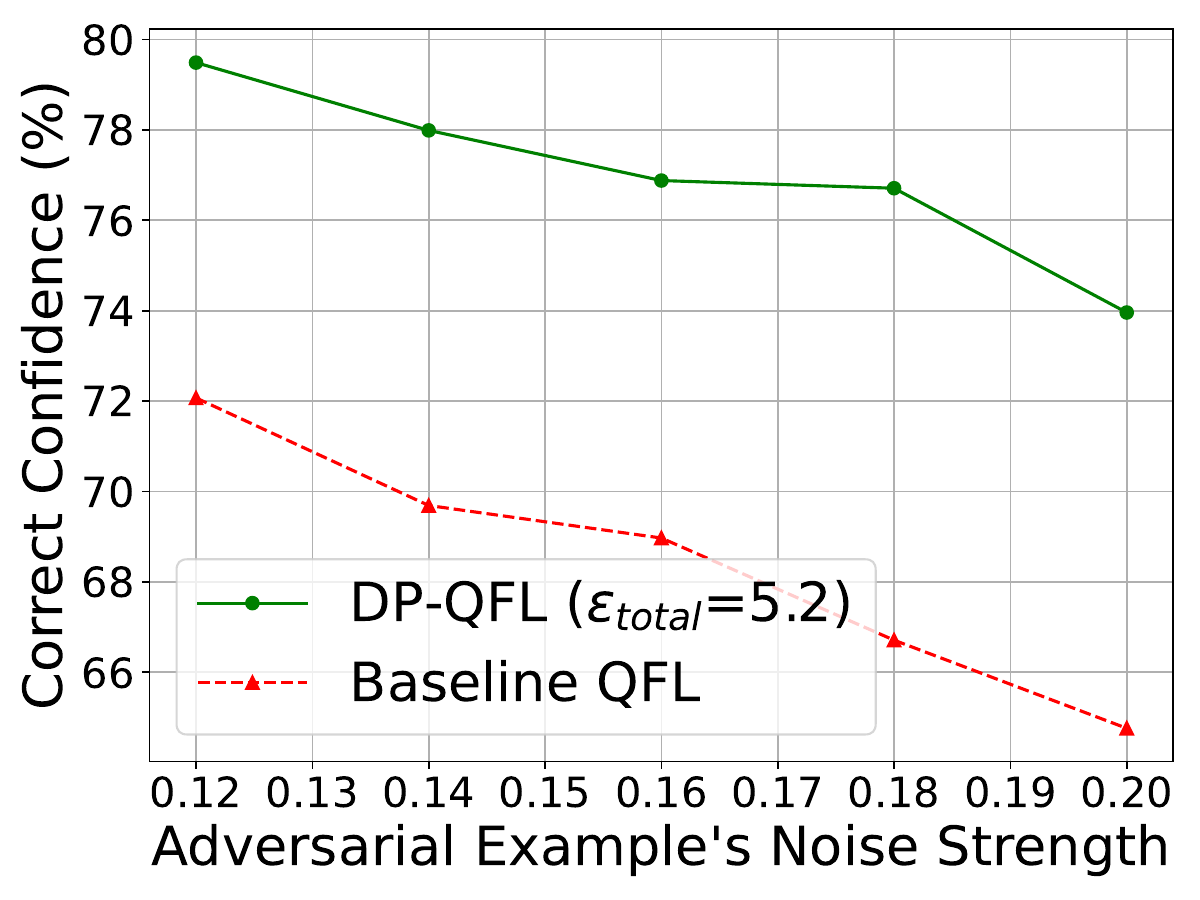}
        \caption{\footnotesize Confidence on correct predictions}
        \label{fig:confidence score}
    \end{subfigure}
    \begin{subfigure}[t]{0.32\linewidth}
        \centering
        \includegraphics[width=\linewidth]{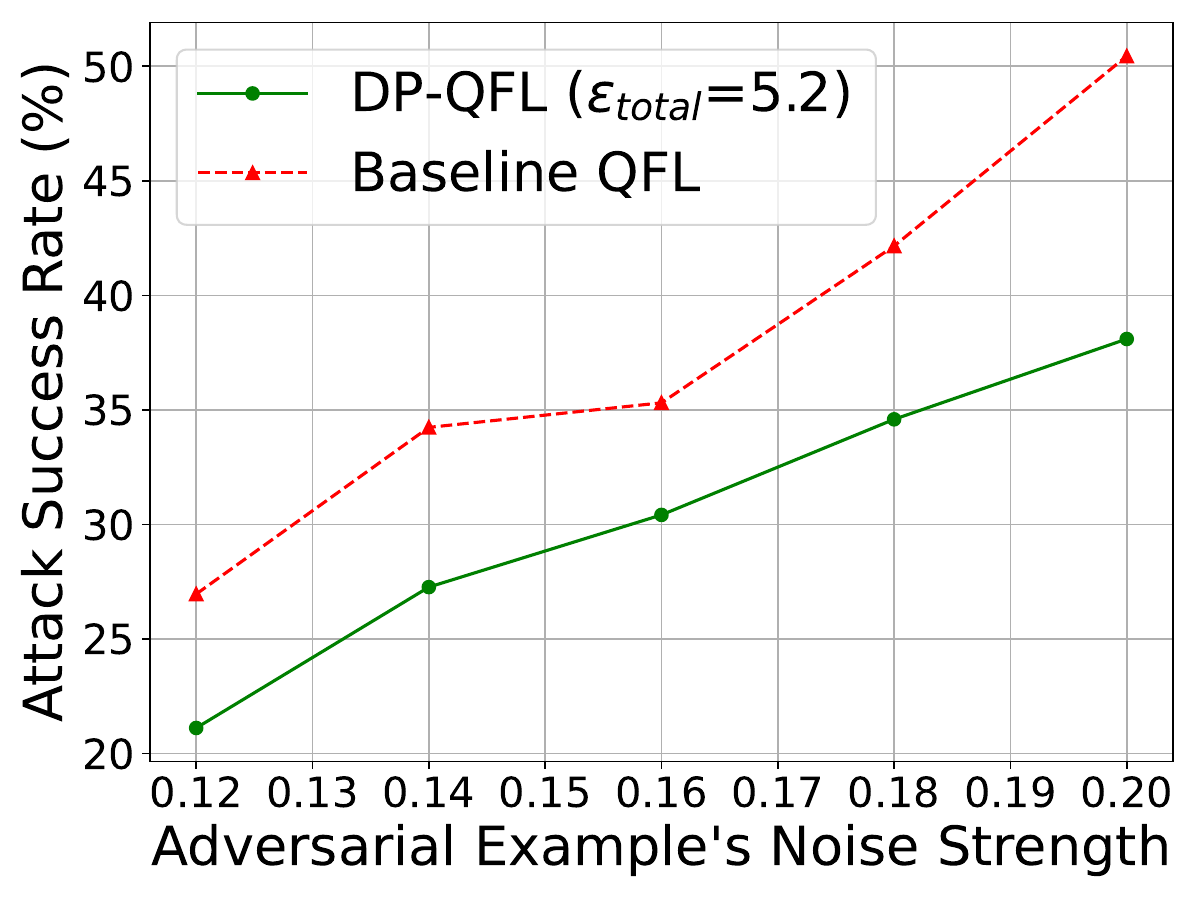}
        \caption{\footnotesize Attack success rate}
        \label{fig:attack_success_rate}
    \end{subfigure}
    
    \vspace{1pt} 
    \caption{Robustness evaluation of DP-QFL under FGSM attack with adversarial examples. The evaluation metrics are: classification accuracy with adversarial examples, the model's confidence in correct predictions, and the attack success rate. }
    \label{fig:qfl_overview}
    \vspace{-2mm}
\end{figure}

Fig. \ref{fig:adversarial accuracy} shows that as adversarial noise strength \( E \) increases from 0.12 to 0.20, both models show reduced accuracy. Throughout this adversarial interval, the DP-protected model consistently performs better, achieving, on average 7\% higher classification accuracy than the non-DP model. In the absence of adversarial noise, the non-DP model achieves 6–8\% higher accuracy than the DP-protected model, reflecting its vulnerability under adversarial conditions. Similarly, the non-DP model exhibits significantly lower confidence in correct predictions than the DP-protected model across all adversarial noise levels, as shown in Fig.~\ref{fig:confidence score}. Subsequently, we compute the attack success rate (ASR) as the ratio of misclassified adversarial examples to the total number of adversarial inputs. The non-DP model shows a 7–20\% higher ASR than the DP-QFL model, as illustrated in Fig. \ref{fig:attack_success_rate}, highlighting the efficacy of DP-QFL against such threats.

\section{Conclusion}
In this study, we presented a novel DP-QFL framework that leverages inherent quantum noise in NISQ devices for differential privacy in federated quantum learning. Our empirical results with two standard datasets demonstrated how parameters such as the number of measurement shots, depolarizing noise factor, and circuit depth affect both the security and performance of the model. To validate the robustness of the framework, we showcased the resilience of DP-QFL against a quantum adversarial attack via multiple metrics. Overall, one novel DP-QFL framework highlighted the feasibility of securely training quantum models in a distributed QFL ecosystem in the NISQ era without requiring additional artificial noise, paving the pathway for practical quantum computing in privacy-critical applications.
    \bibliography{QFL-conference}
\bibliographystyle{IEEEtran}

\end{document}